\documentclass[journal,12pt,draftclsnofoot,onecolumn]{IEEEtran}
\usepackage{mathrsfs}
\usepackage{cite}
\usepackage{amsmath}
\usepackage{amssymb}
\usepackage{stfloats}
\usepackage{graphicx}
\usepackage{setspace}
\usepackage{xcolor}
\usepackage{multirow}
\usepackage{slashbox}

\ifCLASSOPTIONcompsoc
\usepackage[tight,normalsize,sf,SF]{subfigure}
\else
\usepackage[tight,footnotesize]{subfigure}
\fi

\begin{document}
\title{Comparison of OFDM and Single-Carrier for Large-Scale Antenna Systems}

\author{Yinsheng~Liu, Geoffrey Ye Li, and Wei Han.
\thanks{Yinsheng Liu is with the School of Computer Science and Information Technology and State Key Laboratory of Rail Traffic Control and Safety, Beijing Jiaotong University, Beijing 100044, China, e-mail: ys.liu@bjtu.edu.cn.}
\thanks{Geoffrey Ye Li is with ITP lab, school of ECE, Georgia Institute of Technology, Atlanta 30313, Georgia, USA, e-mail: liye@ece.gatech.edu.}
\thanks{Wei Han is with Huawei Technologies, Co. Ltd., China, e-mail:wayne.hanwei@huawei.com.}
}

\maketitle
\doublespacing

\begin{abstract}
\emph{Large-scale antenna} (LSA) or massive \emph{multiple-input multiple-output} (MIMO) has gained a lot of attention due to its potential to significantly improve system throughput. As a natural evolution from traditional MIMO-\emph{orthogonal frequency division multiplexing} (OFDM), LSA has been combined with OFDM to deal with frequency selectivity of wireless channels in most existing works. As an alternative approach, \emph{single-carrier} (SC) has also been proposed for LSA systems due to its low implementation complexity. In this article, a comprehensive comparison between LSA-OFDM and LSA-SC is presented, which is of interest to the waveform design for the next generation wireless systems.
\end{abstract}

\begin{IEEEkeywords}
Large-scale antenna, massive MIMO, OFDM, single-carrier.
\end{IEEEkeywords}

\newpage
\section{Introduction}

\emph{Single-carrier} (SC) has been widely used in many early digital communication systems, including the \emph{second-generation} (2G) cellular systems where data transmission rate is not so high \cite{GLi}. SC is not suitable for high speed data transmission due to the complexity of signal detection at the receiver to address \emph{inter-symbol interference} (ISI) caused by multipath or frequency selectivity of broadband channels. \emph{Orthogonal-frequency-division-multiplexing} (OFDM) can convert a frequency selective channel into a group of flat fading channels to facilitate signal detection at the receiver, and is thus more appropriate for high speed data transmission due to its robustness against frequency selectivity. Furthermore, the complexity of OFDM transceiver can be greatly reduced through \emph{fast Fourier-transform} (FFT). Therefore, OFDM has gradually substituted SC during the last fifteen years and becomes favorable in current wireless systems.\par

\emph{Multiple-input multiple-output} (MIMO) systems have been widely studied during the last two decades due to its ability to improve the capacity and reliability of wireless systems. Since the benefits of MIMO systems can be easily obtained in flat fading channels, MIMO-OFDM has been used as a key technology in many wireless systems such as downlink transmission in \emph{long-term evolution } (LTE).\par

As an extension of traditional MIMO systems, \emph{large-scale antenna} (LSA) or massive MIMO systems have been proposed recently, which can significantly improve system throughput \cite{FRusek}. Through the employment of an excess number of antennas at the \emph{base station} (BS), the channel vectors or matrices between the BS and different users become asymptotically pairwisely orthogonal. In this case, the \emph{matched filter} (MF) asymptotically becomes the optimal detector \cite{JHoydis}. The asymptotical orthogonality among channel vectors or matrices allows multiusers to work in the same bandwidth without \emph{inter-user interference} (IUI) and thus can improve the spectrum efficiency significantly. Furthermore, it has been shown in \cite{HQNgo} that the transmit power for each user is scaled down by the number of antennas or by the square root of the number of antennas, depending on whether accurate \emph{channel state information} (CSI) of an LSA system are available or not, and the energy efficiency can be therefore significantly improved. As the rising of antenna number, the number of RF-chains is also increased. Single-RF technique can be adopted to reduce the number of RF chains in LSA systems \cite{AMohammadi}. \par

Similar to MIMO-OFDM, it is natural to combine LSA with OFDM where the latter is used to convert frequency selective channels into flat fading ones. Therefore, most of existing works in LSA systems consider only flat fading channels since OFDM is presumed \cite{FRusek,JHoydis}. Although straightforward, LSA-OFDM has several drawbacks. On one hand, OFDM has several disadvantages itself. The guard band and \emph{cyclic prefix} (CP) in OFDM cause extra consumption of resources. The \emph{peak-to-average power ratio} (PAPR) of OFDM is also very high, leading to low efficiency of the power amplifier. On the other hand, the MF in traditional implementation of OFDM receiver is conducted for each subcarrier in the frequency domain and therefore the BS needs an FFT module for each receive antenna to convert the received signal into frequency domain, leading to a heavy computation burden. To reduce the complexity, we introduce an MF-OFDM implementation in this article where the MF is performed in the time domain. In this case, the FFT module can be placed after the time-domain MF and thus shared by all antennas, leading to greatly reduced complexity compared to the traditional OFDM.\par

As an alterative approach, SC has been proposed for LSA systems in \cite{APitar,YLiu2}. On one hand, the shortcomings of OFDM can be in general avoided by using SC. For example, SC has been used in LTE uplink to reduce the PAPR, and terminal cost is therefore reduced since it has no need of high performance power amplifier. On the other hand, ISI due to high speed data transmission in SC modulation can be suppressed through a simple MF in LSA systems. In other words, the equalizer in traditional SC receivers can be omitted, and therefore the complexity of SC systems can be greatly reduced. In \cite{APitar}, SC is used for downlink precoding with an MF precoding matrix, which can be extended to uplink according to the duality \cite{DTse}. In \cite{YLiu}, we have considered SC in an LSA system over Rician fading channels. In that case, an equalizer is required at the receiver to suppress multiuser interference caused by the line-of-sight path.\par

Since OFDM and SC have shown their advantages when combined with LSA, both waveforms are advocated in the literature. It is therefore necessary to have a comprehensive comparison of LSA-OFDM and LSA-SC, which is of interest to the design of future wireless networks. The rest of this article is organized as follows. In Section II, we will first introduce LSA systems, and then we will describe LSA-OFDM and LSA-SC in Section III. Their performances are compared in Section IV. Finally, conclusions and technical challenges are discussed in Section V.

\section{LSA Systems}
In an LSA system, the BS is equipped with a large number of antennas, which makes the LSA systems different from the traditional MIMO systems. We will introduce the benefits of LSA systems in the following.

\subsection{Capacity}
The upper bound for the capacity of a point-to-point MIMO system has been derived in \cite{FRusek}. In regular MIMO systems, the capacity depends on the singular values of the channel matrix, and the upper bound cannot be achieved except when the singular values are all equal. When the BS is equipped with a large number of antennas, however, the upper bound can be always achieved asymptotically with the asymptotical orthogonality of channel vectors. The above results also hold for \emph{multiuser MIMO} (MU-MIMO) in an LSA system where the capacity can be achieved with a simple linear MF for both uplink and downlink. \par

If power allocation is taken into account, water-filling algorithm can be used to maximize the channel capacity \cite{AGoldsmith}. In regular MIMO systems, the watering-filling algorithm depends on the small-scaling fading channels. Essentially, the small-scaling effect has been eliminated by using a large number of antennas in LSA systems. However, the large-scale fading effect will still remain. In such a case, different users may have various large-scale fading coefficients, and therefore the power allocation strategy only depends on the large-scale fading effect \cite{HQNgo}.

\subsection{Energy Efficiency}
The energy efficiency can be improved significantly in an LSA system. It has been shown in \cite{HQNgo} that the transmit power can be scaled down by the antenna number if the CSI is perfect. On the other hand, the transmit power is scaled down by the squar root of the antenna number when the CSI is imperfect. In other words, the transmit power can be greatly reduced since the antenna number in an LSA system is very large, and therefore the energy efficiency is significantly improved.\par

Tradeoff between energy efficiency and spectral efficiency in LSA systems has also been studied in \cite{HQNgo}. It has been shown in \cite{HQNgo} that energy efficiency degrades as the increasing of spectral efficiency with perfect CSI. When the CSI is imperfect, energy efficiency and spectral efficiency can increase simultaneously within the low transmit power region while the energy efficiency degrades as the increasing of spectral efficiency within the high transmit power region.\par

\subsection{Signal Detection and Precoding}
Without loss of generality, we will emphasize the signal detection here since the signal detection techniques in uplink can be also used for downlink precoding due to the duality between uplink an downlink \cite{DTse}.\par

When the antenna number is large enough and the channels corresponding to different antennas and different users are independent, the channel vectors for different users are asymptotically orthogonal \cite{FRusek}. In such a case, the typical linear receivers, such as the \emph{zero-forcing} (ZF) receiver or the \emph{minimum mean-square-error} (MMSE) receiver\cite{DTse}, reduce to a simple MF receiver. Therefore, the receiver complexity is greatly reduced since no matrix inversion is required. Besides, the optimality of the MF receiver has also been proved in \cite{HQNgo} where it is shown that the capacity of a point-to-point MIMO or an MU-MIMO can be achieved asymptotically with the MF when the antenna number is large enough. \par

In general, the asymptotical orthogonality only holds in the ideal case where the antenna number is infinite. In practical systems, however, the antenna number is always finite and thus the MF will not perform as well as expected, especially when the user number is relatively large. It has been shown in \cite{JHoydis} that ZF and MMSE precodings can significantly outperform the MF precoding at the cost of increased complexity. Truncated polynomial expansion based on Cayley-Hamilton theorem has been adopted to reduce the complexity of ZF or MMSE precoding where the performance can be improved iteratively by increasing the expansion order \cite{AKammoun}.\par

In addition to the advantages above, the benefits of LSA systems also include low-complexity scheduling in a multicell environment by exploiting the large number of antennas \cite{HHuh} and so on. Although LSA has shown many advantages over existing systems, a number of issues have to be addressed before it can be applied in practice. The main ones among those issues includes pilot contamination, overhead of feedback, and waveform design.\par

Ideally, the pilot sequences of different users for uplink are expected to be orthogonal so that the BS can obtain the channels of different users without interference. In practical systems, however, the number of orthogonal pilot sequences is limited for a given training period and subcarriers. Therefore, non-orthogonal pilot sequences have to be adopted when the user number is larger than that of orthogonal pilot sequences, leading to pilot contamination \cite{JJose2}. This is also the case for the downlink channel estimation since the number of antenna is very large in LSA systems. The pilot sequences will consume too much resources if orthogonal pilot sequences are used for each antenna at the BS.\par

\emph{Time-division duplexing} (TDD) has been assumed in most existing works where channel reciprocity can be exploited to obtain CSI at the BS \cite{JHoydis,FRusek}. In practical systems, however, \emph{frequency-division duplexing} (FDD) is more favorable where the CSIs are quantized and fed back to the BS through limited feedback technique. This approach works well in regular MIMO systems where the antenna number is small. In LSA systems, however, the antenna number is greatly increased, leading to a huge feedback overhead. A two-stage precoding technique has been developed to reduce the feedback overhead where the inner precoder is adaptive to the instantaneous effective CSI while the outer precoder is only adaptive to the channel statistics \cite{ALiu}. In this way, the feedback overhead can be greatly reduced since the inner precoder has a lower dimension.\par

For waveform design, OFDM has been presumed in most LSA systems, which can be considered as a natural evolution from traditional MIMO-OFDM. Although straightforward, several shortcomings of LSA-OFDM cannot be avoided. As an alternative, SC has also been adopted in LSA systems to mitigate the disadvantages of OFDM \cite{APitar,YLiu2}. Note that besides OFDM and SC, there are also other flexible waveforms that are under investigation for the next generation wireless systems, such as \emph{filter-bank multicarrier} (FBMC) and \emph{universal filtered multicarrier} (UFMC) \cite{FSchaich}. Although novel, the characteristics of those waveforms are still under research and they have not been utilized in practical systems. On the other hand, OFDM and SC have been well understood and accepted during the past decades. As basic waveforms, the reliabilities of OFDM and SC have been verified in many existing systems. Considering the reliability and the compatibility with existing systems, we only focus OFDM and SC in this article while the other waveforms are not taken into account.\par

We refer the readers to the literature above for more information about pilot contamination and overhead of feedback. In the subsequential of this article, we will focus on the issue of waveform design. In particular, a detailed comparison between LSA-OFDM and LSA-SC will be presented.

\section{Waveform Design}
In this section, we will first introduce the channel model in LSA systems. Then, we will present LSA-OFDM and LSA-SC, respectively.
\subsection{Channel Model}

For uplink transmission in an LSA system with $M$ receive antennas, the basedband CIR vector from the user to the BS can be given by
\begin{align}\label{e1}
\mathbf{h}(t)=\sum_{l}\mathbf{c}[l]\delta(t-\tau_l),
\end{align}
where $\mathbf{h}(t)=[h_1(t),\cdots,h_M(t)]^{\mathrm{T}}$ with $h_m(t)$ denoting the CIR at the $m$-th receive antenna, $\mathbf{c}[l]=(c_1[l],\cdots,c_M[l])^{\mathrm{T}}$ with $c_m[l]$ denoting the complex gain of the $l$-th tap, $\delta(\cdot)$ is the Dirac-delta function, and $\tau_l$ denotes the corresponding tap delay. The extra propagation delays caused by the physical size of the antenna array are omitted since they are quite small, and thus different antennas have the same tap delay in (\ref{e1}). From (\ref{e1}), the \emph{channel frequency response} (CFR) vector, $\mathbf{H}(f)=[H_1(f),H_2(f),\cdots,H_M(f)]^{\mathrm{T}}$, can be expressed as
\begin{align}\label{e2}
\mathbf{H}(f)=\sum_l\mathbf{c}[l]e^{-j2\pi f\tau_l},
\end{align}
where $H_m(f)$ denotes the CFR at the $m$-th receive antenna.\par

Following the \emph{wide-sense-stationary uncorrelated-scattering} (WSSUS) assumption in \cite{AGoldsmith} the complex gains corresponding to different taps are independent. As a result, if the antenna number in an LSA system is large enough, then the complex gain vectors for different taps are asymptotically orthogonal, that is
\begin{align}\label{e3}
\frac{1}{M}\mathbf{c}^{\mathrm{H}}[l]\mathbf{c}[l_1]\rightarrow\sigma_{l}^2\delta[l-l_1],~\text{as}~M\rightarrow\infty,
\end{align}
where $\sigma_{l}^2$ is the power of the $l$-th tap. Using the asymptotical orthogonality in (\ref{e3}), a simple MF detector is enough to mitigate ISI in SC systems \cite{APitar,YLiu2}. The asymptotical orthogonality can be also used in the frequency domain. In this case, the CFR vectors corresponding to different users will be asymptotically orthogonal when the antenna number is large enough, and thus the IUI can be easily canceled through an MF \cite{FRusek}. Similar to the CFR vector case, the CIR vectors in (\ref{e1}) corresponding to different users are also asymptotically orthogonal and thus the IUI can be also mitigated easily. Therefore, in the subsequent discussion, we assume that there is only one user and no IUI occurs even if there may be multiple users in the system.\par

Note that we only consider the uplink transmission in this article. According to the duality \cite{DTse}, the results in this article can be also used for the downlink if the corresponding CSIs are known at the BS.\par

\subsection{LSA-OFDM}

\begin{figure*}
  \centering
  \includegraphics[width=6in]{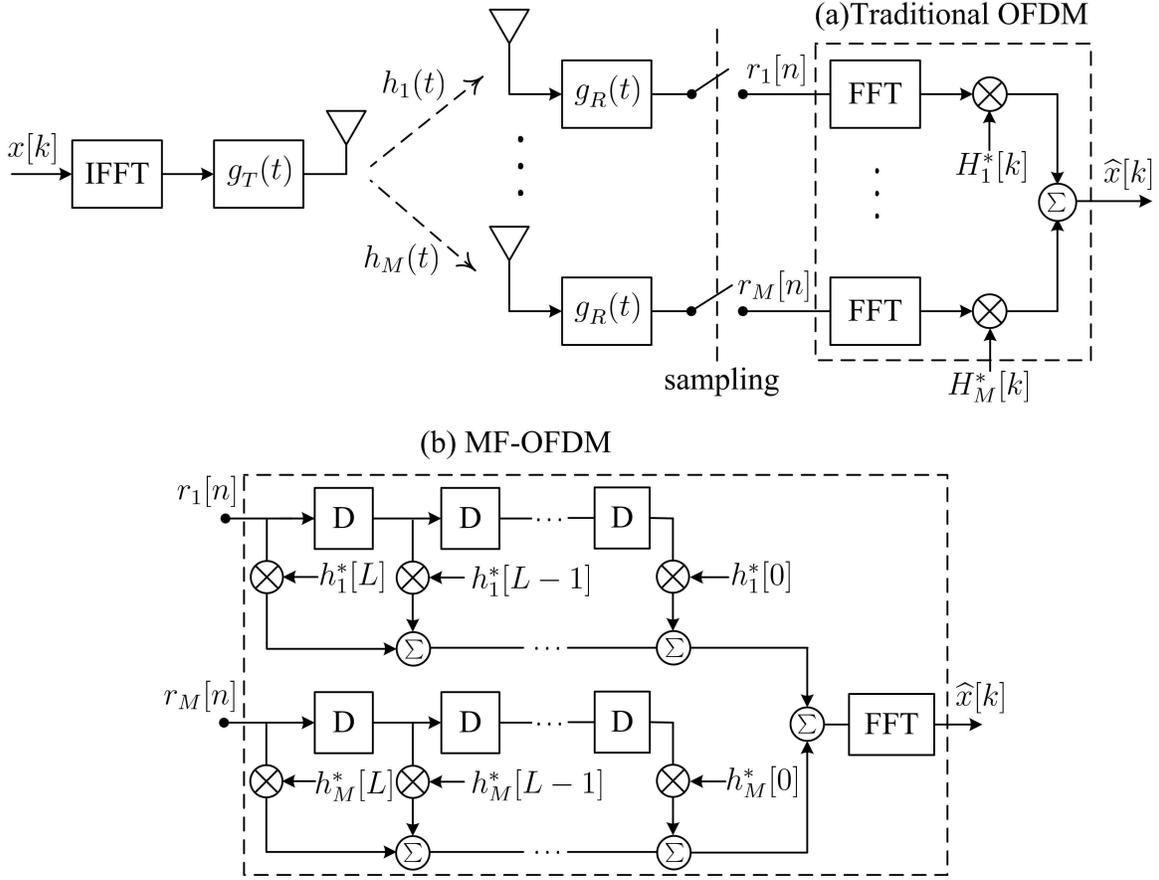}\\
  \caption{LSA-OFDM systems with (a) traditional OFDM implementation and (b) MF-OFDM implementation.}\label{LSA_OFDM}
\end{figure*}
In regular MIMO systems, OFDM is used to convert a frequency-selective fading channel into a group of flat fading subchannels. As a natural evolution of MIMO-OFDM, LSA systems have used OFDM in most existing works, such as in \cite{FRusek,HQNgo}.\par

For an LSA-OFDM system based on traditional OFDM implementation, the receiver structure is shown in Fig.~\ref{LSA_OFDM} (a), where $x[k]$ denotes the transmit symbol on the $k$-th subcarrier, $r_m[n]$ denotes the $n$-th sample of the received signal at the $m$-th antenna, and $g_T^{}(t)$ and $g_R^{}(t)$ are the transmit and receive filters, respectively. The transmit and receive filters are used to model the effect of analog front-end, and thus they are required by both OFDM and SC, as in \cite{GLi}. In the figure, $H_m[k]=H_m(k\Delta f)$ denotes the CFR on the $k$-th subcarrier at the $m$-th antenna, where $\Delta f$ is the subcarrier spacing in OFDM. Using the asymptotical orthogonality property, MF is enough for optimal signal detection in an LSA-OFDM system, which is different from the traditional MIMO-OFDM. Note that the MF for the traditional implementation of OFDM is conducted for each subcarrier in the frequency domain and therefore each receive antenna needs an FFT module to convert the received signal into frequency domain.\par

Since the multiplication in the frequency domain corresponds to the circular convolution in the time domain, the traditional OFDM can be also implemented equivalently using the MF-OFDM structure as in Fig.~\ref{LSA_OFDM} (b), where $h_m[l]$ with $l=0,1,\cdots, L$ denotes the sampled CIR. Essentially, $h_m[l]$ is exactly the inverse discrete Fourier transform of $H_m[k]$. Due to the correlation of CFR, the sampled CIR length, $L+1$, is in general much smaller than the FFT size. Note that the received signal should be circularly extended before sending to the time-domain MF so that the received signal can be circularly convolved with the time-domain MF, as expected by the property of the DFT. In this case, the MF is conducted in the time domain before the FFT module, and therefore the complexity can be greatly reduced since all antennas share only one FFT module.

\subsection{LSA-SC}

\begin{figure*}
  \centering
  \includegraphics[width=5.5in]{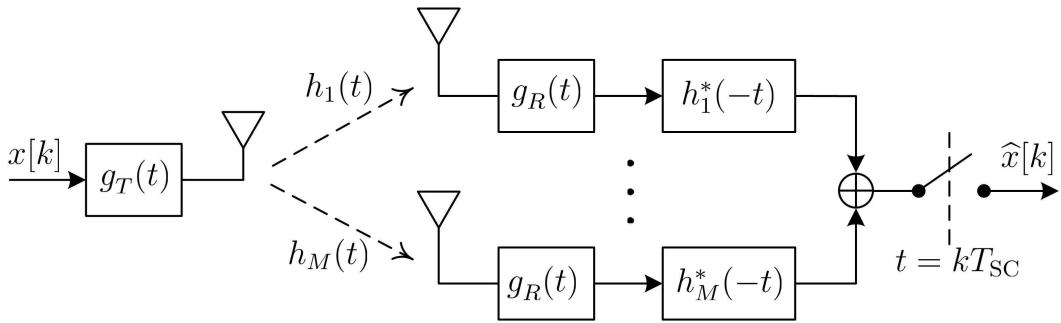}\\
  \caption{System structure for an LSA-SC System.}\label{LSA_SC}
\end{figure*}

SC has also been proposed for an LSA-SC system \cite{APitar,YLiu2} as in Fig.~\ref{LSA_SC}, where $x[k]$ denotes the transmit symbol at the $k$-th time instance and $T_{\mathrm{SC}}$ denotes the symbol duration for SC modulation. The LSA-SC receiver in Fig.~\ref{LSA_SC} can be derived as a simplification of the traditional SC receiver with multiple receive antennas \cite{GLi}. Following the analog MF, an equalizer is required after the baudrate sampling in traditional SC receivers. When the antenna number is large enough, the equalizer reduces to a Kronecker delta function and thus can be omitted from the receiver. In brief, this is similar to the fact that ZF or MMSE receiver in regular MIMO systems will reduce to the MF receiver in LSA systems. Similarly, the ZF or MMSE equalizer for ISI suppression in traditional SC systems also reduces to an MF in LSA-SC systems.\par

Alternatively, the receiver structure in LSA-SC systems can be also explained through waveform recovery theory \cite{YLiu2}. Using the asymptotical orthogonality property in (\ref{e2}), the tap vectors, $\mathbf{c}[l_1]$ and $\mathbf{c}[l_2]$, are asymptotically orthogonal for $l_1\neq l_2$ when the antenna number is large enough. In this case, the waveform transmitted over each tap can be easily extracted using a single-tap MF receiver. To improve the receiver performance, the extracted waveforms transmitted over all taps can be combined and therefore the resulted receiver structure is the same with that in Fig.~\ref{LSA_SC}.\par

Similar to MF-OFDM, the MF for LSA-SC is also conducted in the time domain except that the MF is an analog one. Essentially, the receiver structure in Fig.~\ref{LSA_SC} is only a theoretical result. It is in general hard to implement since the CIR, $h_m(t)$, is analog and it is therefore difficult to obtain the estimation of such an analog CIR. Even though the analog CIR is known in advance, it is still not implementable because the analog front-end is, in general, fixed in the receiver and we are thus unable to adjust it adaptively \cite{AGoldsmith}. For practical receiver designs, oversampling based receiver has been advocated in the existing works \cite{GLi,YLiu2}, where the analog CIR can be implemented equivalently with sampled CIR and thus can be obtained through channel estimation techniques.\par

\section{Comparisons of LSA-OFDM and LSA-SC}
In this section, we will compare LSA-OFDM and LSA-SC in terms of implementation complexity, \emph{block-error rate} (BLER), out-of-band emission, spectral efficiency, energy efficiency, and effects of different antenna numbers, respectively.\par
Consider an OFDM-based LTE systems, where the subcarrier spacing is $\Delta f=15~\text{KHz}$ corresponding to an OFDM symbol duration of $T_{\text{OFDM}}=1/{\Delta f}\approx 66.7 \mu\text{s}$ and the CP duration is about $7\%$ of the OFDM symbol duration. For a typical $5$ MHz bandwidth in LTE, the size of FFT for OFDM is $N_{\mathrm{FFT}}=512$, among which $N=300$ subcarriers are used for data transmission and the others are used as guard band. In the presence of the CP and the guard band, the OFDM can transmit about $4.2\times 10^6$ symbols per second. For fair comparison, the symbol rate for SC is the same as that of OFDM, corresponding to a SC symbol duration of $T_{\text{SC}}\approx 0.24 \mu\text{s}$. \emph{Root raised-cosine }(RRC) filters with roll-off factor $\beta=0.22$ are used for transmit filtering and receive filtering in both SC and OFDM.\par

The same modulation coding schemes are used for both LSA-OFDM and LSA-SC, including {\emph{quadrature-phase-shift-keying}} (QPSK), \emph{16-point quadrature-amplitude-modulation} (16QAM) and unpunctured LTE standardized turbo code ($1/3$ code rate) with an information block length of $614$ bits. We consider a normalized large-scale fading in this article, and an \emph{extended typical urban} (ETU) channel model is used for small-scale fading, which has a maximum delay $\tau_{\mathrm{max}}=5\mu\text{s}$. Unless specified, we consider $M=100$ antennas at the BS and channels corresponding to different antennas are assumed to be independent.\par

\subsection{Implementation Complexity}
\begin{table*}
  \centering
  \caption{Comparison of complexity for LSA-OFDM and LSA-SC.}\label{tab_complexity}
  \begin{tabular}{|c||c|c|c|c|c|}
  \hline
  \multirow{2}{*}{\backslashbox{Waveform}{Complexity}} & \multirow{2}{*}{$M=2$} &  \multirow{2}{*}{$M=8$} & \multirow{2}{*}{$M=32$} & \multirow{2}{*}{$M=128$} & \multirow{2}{*}{$M=512$}\\
  & & & & & \\
  \hline
  \multirow{2}{*}{LSA with traditional OFDM} &  \multirow{2}{*}{$5.2\times 10^3$} & \multirow{2}{*}{$2.1\times 10^4$} & \multirow{2}{*}{$8.3\times 10^4$} & \multirow{2}{*}{$3.3\times 10^5$} & \multirow{2}{*}{$1.3\times 10^6$}\\
  & & & & & \\
  \hline
  \multirow{2}{*}{LSA with MF-OFDM} & \multirow{2}{*}{$2.4\times 10^3$}  &  \multirow{2}{*}{$2.6\times 10^3$} & \multirow{2}{*}{$3.6\times 10^3$} & \multirow{2}{*}{$7.3\times 10^3$} & \multirow{2}{*}{$2.2\times 10^4$}\\
  & & &  & & \\
  \hline
  \multirow{2}{*}{LSA-SC} & \multirow{2}{*}{$0.2\times 10^3$}  & \multirow{2}{*}{$0.6\times 10^3$} & \multirow{2}{*}{$2.5\times 10^3$} & \multirow{2}{*}{$9.9\times 10^3$} & \multirow{2}{*}{$3.9\times 10^4$}\\
  & & &  & & \\
  \hline
\end{tabular}
\end{table*}
The implementation complexity is evaluated in terms of the numbers of multiplications.\par
LSA-OFDM receiver based on the traditional OFDM implementation needs $\frac{N_{\mathrm{FFT}}}{2}\log_2N_{\mathrm{FFT}}$ multiplications for each FFT and $N$ multiplications for the MF on each antenna. Therefore, the total number of multiplications required is $M({\frac{N_{\mathrm{FFT}}}{2}}\log_2N_{\mathrm{FFT}}+N)$. For the MF-OFDM, only one FFT is required in the receiver. Hence, the total number of multiplications is $M(L+1)+\frac{N_{\mathrm{FFT}}}{2}\log_2 N_{\mathrm{FFT}}$, which is much fewer than that of the traditional OFDM implementation..\par

For the oversampling based implementation of the LSA-SC receiver, the oversampling rate is defined as $\alpha=f_sT_{\text{SC}}$, where $f_s$ is the sample frequency of the SC signal. In such circumstance, the LSA-SC receiver needs $\alpha (L+1)$ multiplications for the MF at each antenna. Therefore, the number of overall multiplications required for LSA-SC is $M(L+1)\alpha$.\par

We consider the $5$ MHz bandwidth case in LTE as an example. Given the sampling frequency of $7.68 \mathrm{MHz}$ and the maximum delay spread of $5\mu$s for an ETU channel, a sampled CIR length of $L=38$ is enough to include most of the channel power. For LSA-SC, the oversampling rate is usually an integer. It is shown in \cite{GLi} that a twice oversampling rate with $\alpha=2$ can already achieve satisfied performance. In this circumstance, the complexities of LSA-SC and LSA-OFDM based on traditional OFDM and MF-OFDM with respect to different antenna numbers can be shown in Tab.~\ref{tab_complexity}. When the antenna number is small, all the approaches have similar complexities. Compared to traditional OFDM, the complexity of LSA-SC reduces two orders of magnitude when the antenna number is large enough while its complexity is similar to that of the MF-OFDM. Such observation is not surprising. The traditional OFDM implementation requires an FFT at each antenna, leading to huge computation burden. For LSA-SC and MF-OFDM, the MF is conducted in the time domain and thus can achieve lower complexity.\par

\subsection{BLER}
\begin{figure}
  \centering
  \includegraphics[width=5in]{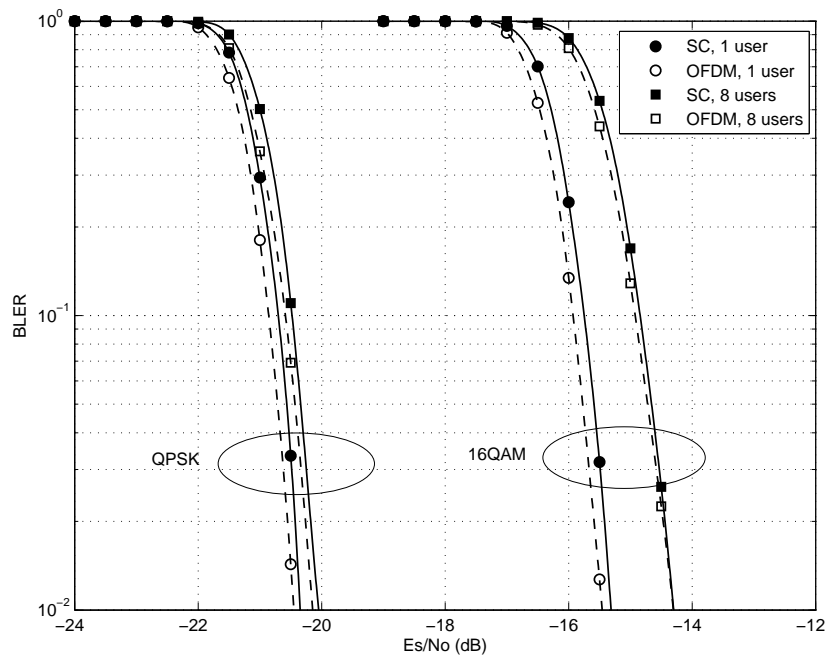}\\
  \caption{BLERs for LSA-SC and LSA-OFDM.}\label{fig_BLER}
\end{figure}

Since the traditional OFDM implementation and the MF-OFDM implementation are functionally equivalent, we therefore do not distinguish them in the following discussion. Fig.~\ref{fig_BLER} shows the BLER versus \emph{signal-to-noise ratio} (SNR), which is defined as $E_s/N_0$ with $E_s$ being the energy for the modulation symbol and $N_0$ being the power spectrum density of the additive white Gaussian noise. As expected, we can observe a water-fall region in the figure since the small-scale fading has been eliminated. In brief, we have the following observations from the figure.
\subsubsection{Residual ISI}
For the single user case, LSA-OFDM outperforms LSA-SC by about $0.2$ dB at $\text{BLER}=10^{-2}$. As a natural evolution from MIMO-OFDM, the frequency selectivity in LSA-OFDM systems can be always addressed through OFDM, and thus there is no residual ISI. For LSA-SC, the ISI is suppressed using the asymptotical orthogonality in LSA systems. The ISI can be completely removed only if the antenna number is theoretically infinite. In practical systems, however, the antenna number is always finite. This leads to irreducible residual ISI in LSA-SC systems, and thus LSA-OFDM outperforms LSA-SC. Even though, the interference power can be very small when the antenna number is large \cite{YLiu2}. Therefore, the performance degradation is quite small and it can be easily compensated for by increasing the transmit power for only about $0.2$ dB.
\subsubsection{Residual IUI}
Both LSA-OFDM and LSA-SC have performance degradations in the presence of multiple users. For both LSA-OFDM and LSA-SC, the IUI is suppressed using the asymptotical orthogonality in LSA systems. Since the antenna number is finite, the IUI cannot be canceled completely, resulting in irreducible residual IUI. Therefore, the performance gets worse as the number of users increases. As also observed, the BLER for higher order modulation, such as 16QAM, is more sensitive to the increasing of user number since the higher order modulation has more dense constellation given the same symbol energy and thus be more apt to be affected by the residual IUI. Nevertheless, the performance gap between LSA-OFDM and LSA-SC gets smaller or even vanishes for 16QAM in the multiuser case because the residual IUI rather than the residual ISI is dominant in the case of multiusers and thus both LSA-OFDM and LSA-SC can achieve the same performance.

\begin{figure}
  \centering
  \includegraphics[width=5in]{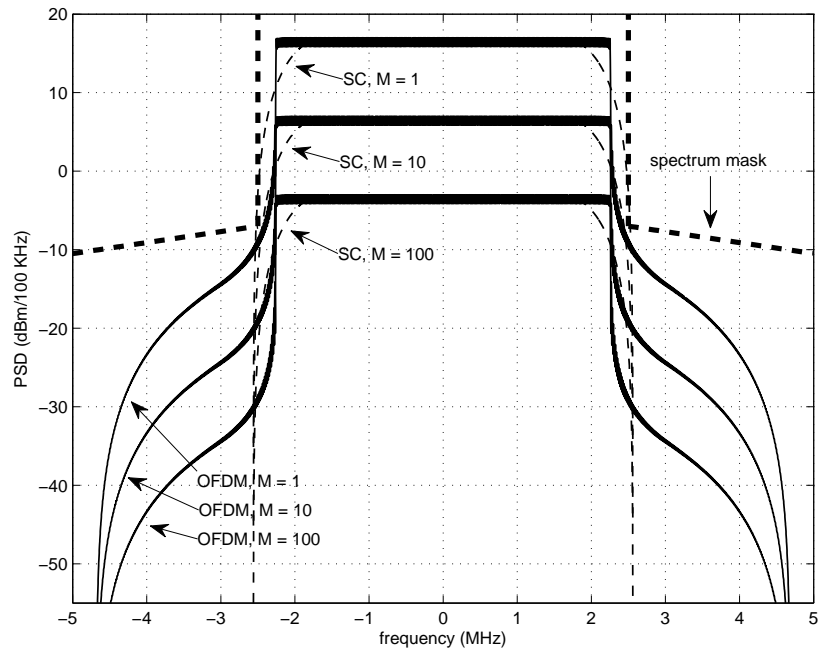}\\
  \caption{Spectrum structure for OFDM and SC.}\label{fig_spectrum}
\end{figure}

\subsection{Out-of-Band Emission}
The out-of-band emission depends on the spectrum structure. In theory, the spectrum structure is only determined by the shape of the transmit filter. In practice, however, it is also affected by the truncation length of the transmit filter and filter order. Besides, the resolution bandwidth of the measurement device also affects the observed spectrum. Although important, it makes the evaluation of the spectrum structure intractable if taking these engineering factors into account. Therefore, to gain insights on the comparison between LSA-OFDM and LSA-SC, those engineering factors are ignored here, that is, we assume the transmit filter can be implemented ideally and the measurement device has a perfect resolution.\par

Fig.~\ref{fig_spectrum} shows the \emph{power spectrum density} (PSD) for OFDM and SC. From the figure, the spectrum mask for existing LTE systems can be satisfied more easily by increasing the antenna number because the transmit power is greatly reduced due to the large array gain of LSA systems. On the other hand, out-of-band emission of OFDM is stronger than that of SC since a rectangular impulse is implicitly used for OFDM, leading to large out-of-band emission. Therefore, OFDM will cause more severe adjacent channel interference than SC.\par

\begin{figure}
  \centering
  \includegraphics[width=5in]{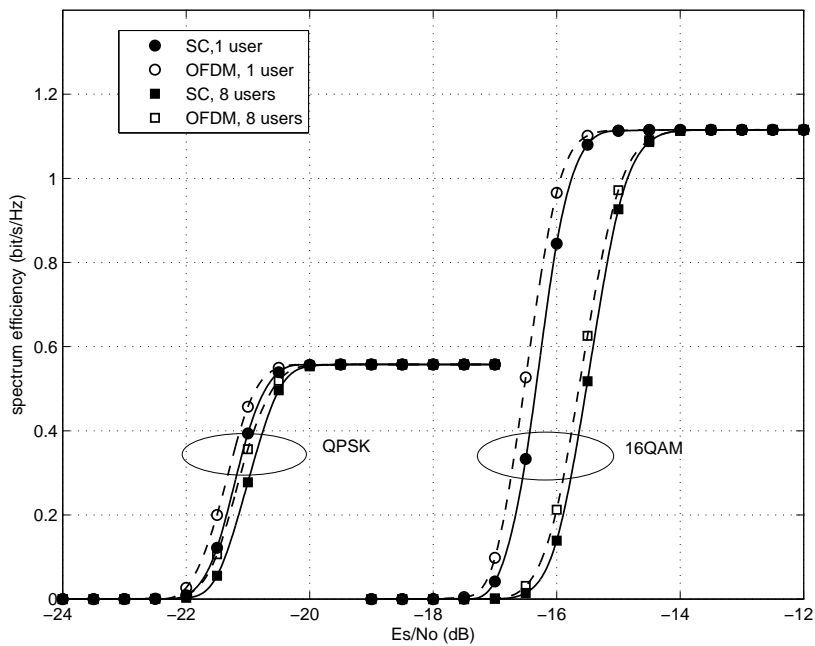}\\
  \caption{Spectrum efficiency for both LSA-OFDM and LSA-SC.}\label{fig_SE}
\end{figure}

\subsection{Spectrum Efficiency}

The spectrum efficiency, in units of bit/s/Hz, is defined as the information rate that can be transmitted over a given bandwidth. Since SC have the same symbol rate with traditional OFDM, they can achieve the same spectrum efficiency when SNR is large enough, as shown in Fig.~\ref{fig_SE}. Similar to the BLER case, we can also observe an about $0.2$ dB performance gap between LSA-OFDM and LSA-SC in Fig.~\ref{fig_SE}. This is because more symbol energy is needed for LSA-SC to cancel out the impact of residual ISI. In the presence of multiple users, both LSA-OFDM and LSA-SC have performance degradations due to the similar reason as the BLER. Also, the higher order modulation is more sensitive to the increase of user number. Besides, the performance gap between LSA-OFDM and LSA-SC for 16QAM vanishes when SNR is large enough and the reason is also the same with that for the BLER. Note that the spectrum efficiency is averaged over all users, and thus the results of multiuser case are the same with that of the single-user case.\par

\subsection{Energy Efficiency}

\begin{figure}
  \centering
  \includegraphics[width=5in]{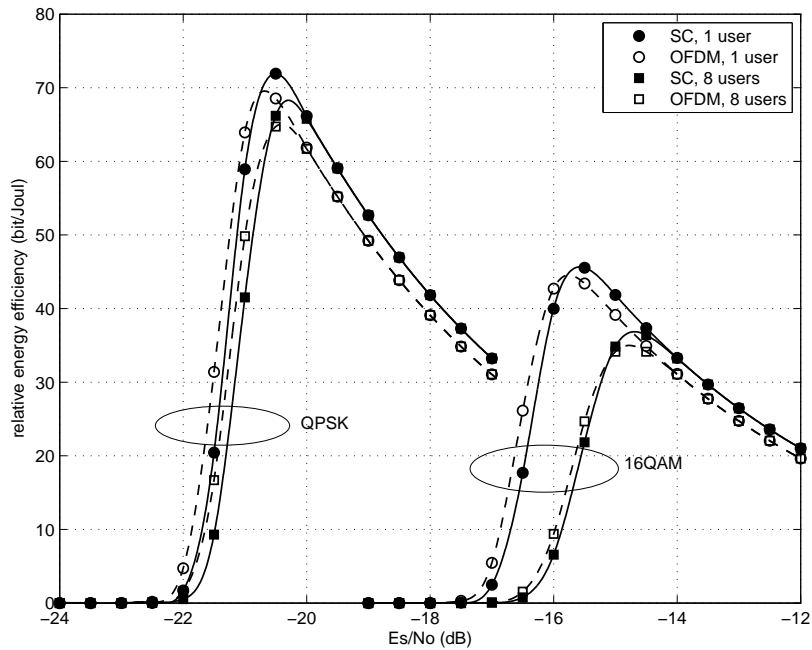}\\
  \caption{Relative energy efficiencies for LSA-OFDM and LSA-SC.}\label{fig_EE}
\end{figure}
Energy efficiency, in units of bits/Joul, is defined as the information rate that can be transmitted for a given transmit power. In general, the effect of large-scale fading should be taken into account to derive a true energy efficiency. Since the large-scale effect is normalized in this article, the derived results are relative energy efficiencies \cite{HQNgo}.\par

Fig.~\ref{fig_EE} compares the relative energy efficiencies for both LSA-OFDM and LSA-SC. From the figure, the energy efficiency of LSA-OFDM outperforms that of LSA-SC when SNR is small because LSA-SC needs more symbol energy to cancel out the impact of residual ISI. When SNR is large, both LSA-OFDM and LSA-SC achieve the maximum rate and thus the impact of the residual ISI vanishes. In this case, LSA-SC shows better energy efficiency performance because LSA-OFDM has to consume extra power for the transmission of the CP. Similar to the above situation, the higher order modulation is more sensitive to the increase of user number. We also note that the performance for single-user and multiuser cases is the same when SNR is large enough. In this situation, both single-user and multiuser cases can achieve the same transmission rate as shown in Fig.~\ref{fig_SE}, and thus the energy efficiency is also the same given the same symbol energy.

\subsection{Effect of Antenna Number}
\begin{figure}
  \centering
  \includegraphics[width=5in]{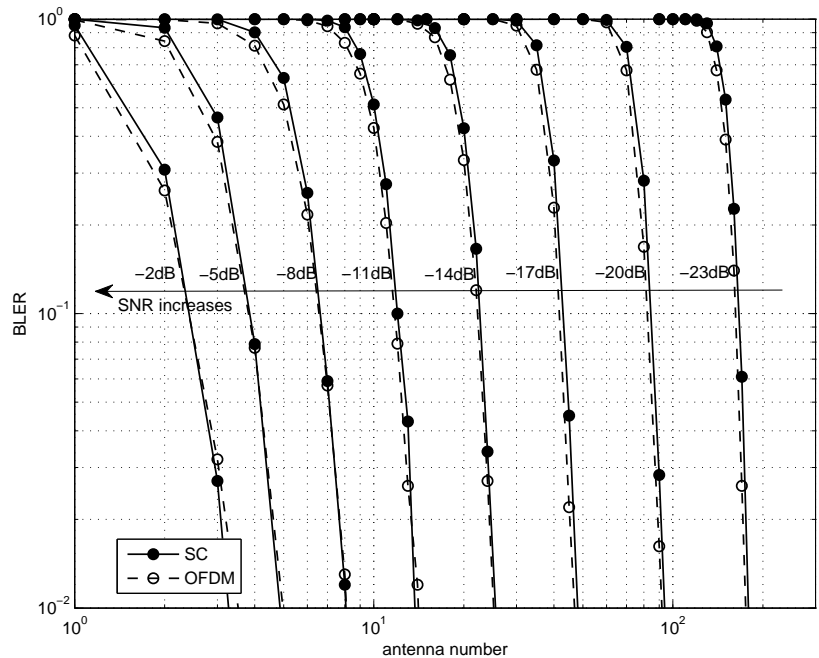}\\
  \caption{BLERs with different numbers of antennas.}\label{fig_BLER_ant}
\end{figure}

In the above, we only consider the case of $M=100$. Fig.~\ref{fig_BLER_ant} shows the BLERs for different numbers of antennas where the QPSK modulation is used for a single-user scenario. From the figure, LSA-SC needs about $3\%$ more antennas to achieve the same performance with LSA-OFDM since the former suffers from residual ISI. As the antenna number increases, similar to Fig.~\ref{fig_BLER}, the curves also show a water-fall region where the BLER drops sharply when the antenna number is larger than a threshold determined by the SNR. In other words, for a given SNR, there is no need to employ more antennas when the number of antenna exceeds that threshold since the BLER has been good enough (BLER$ < 10^{-2}$ corresponds to less than $1\%$ loss of the spectrum efficiency). Fig.~\ref{fig_BLER_ant} also shows that large number of antennas can be used to compensate for the low transmit power when the SNR is small. When the SNR is large, small antenna numbers are enough to achieve satisfied performance and large number of antennas are therefore not necessary in this case.

\section{Discussions and Conclusions}
In this article, we have compared LSA systems combined with OFDM and SC. From our discussion, the following conclusions can be obtained.
\begin{itemize}
\item {LSA-SC can achieve a lower complexity than LSA-OFDM based on the traditional OFDM, and it has a similar complexity with the MF-OFDM. }

\item{When SNR is large, LSA-OFDM and LSA-SC can achieve the same spectral efficiency, while LSA-SC has an about $0.2$ dB degradation compared to LSA-OFDM when SNR is small.}
\item{The energy efficiency of LSA-SC is higher than that of LSA-OFDM when SNR is large. When SNR is small, the energy efficiency of LSA-SC is lower than that of LSA-OFDM.}

\item{Compared to LSA-OFDM, LSA-SC suffers from an $0.2$ dB degradation for BLER at $\mathrm{BLER}=10^{-2}$ also due to residual ISI.}
\item{Existing spectrum mask can be easily satisfied for both LSA-SC and LSA-OFDM since the transmit power is greatly reduced. Given the same symbol rate, the out-of-band emission of SC is much smaller than OFDM, and therefore SC causes much smaller adjacent channel interference than that of OFDM.}
\item{The performance degradation caused by residual ISI in LSA-SC can be compensated for by using $3\%$ more antennas.}
\end{itemize}\par
In the above, we have compared different aspects of LSA-OFDM and LSA-SC. If taking into account the evolution from existing systems to future wireless networks, LSA-OFDM based on MF-OFDM can be compatible with current frame structure of the transmit signal and thus allows for a smooth evolution from regular MIMO to LSA with lower complexity.\par
Our result in this article is also applicable for downlink scenario if the downlink channel is known. However, such channel knowledge is usually unknown at the BS. Therefore, efficient CSI feedback technique or reciprocity based TDD technique should be developed for the practical applications.

\bibliographystyle{IEEEtran}
\bibliography{IEEEabrv,lsmimobib}

\begin{thebibliography}{10}
\providecommand{\url}[1]{#1}
\csname url@samestyle\endcsname
\providecommand{\newblock}{\relax}
\providecommand{\bibinfo}[2]{#2}
\providecommand{\BIBentrySTDinterwordspacing}{\spaceskip=0pt\relax}
\providecommand{\BIBentryALTinterwordstretchfactor}{4}
\providecommand{\BIBentryALTinterwordspacing}{\spaceskip=\fontdimen2\font plus
\BIBentryALTinterwordstretchfactor\fontdimen3\font minus
  \fontdimen4\font\relax}
\providecommand{\BIBforeignlanguage}[2]{{%
\expandafter\ifx\csname l@#1\endcsname\relax
\typeout{** WARNING: IEEEtran.bst: No hyphenation pattern has been}%
\typeout{** loaded for the language `#1'. Using the pattern for}%
\typeout{** the default language instead.}%
\else
\language=\csname l@#1\endcsname
\fi
#2}}
\providecommand{\BIBdecl}{\relax}
\BIBdecl

\bibitem{GLi}
G.~Y. Li, J.~H. Winters, and N.~R. Sollenberger, ``Spatial-temporal
  equalization for {IS}-136 {TDMA} systems with rapid dispersive fading and
  cochannel interference,'' \emph{IEEE Trans. Veh. Techno.}, vol.~48, no.~4,
  pp. 1182--1194, July 1999.

\bibitem{FRusek}
F.~Rusek, D.~Perrsson, B.~K. Lau, E.~G. Larsson, T.~L. Marzetta, O.~Edfors, and
  F.~Tufvesson, ``Scaling up {MIMO}: opportunities and challenges with very
  large arrays,'' \emph{IEEE Signal Process. Mag.}, vol.~30, no.~1, pp. 40--60,
  Jan. 2013.

\bibitem{JHoydis}
J.~Hoyidis, S.~ten Brink, and M.~Debbah, ``Massive {MIMO} in {UL}/{DL} of
  cellular networks: How many antennas do we need?'' \emph{IEEE J. Sel. Area
  Commun.}, vol.~31, no.~2, pp. 160--171, Jan. 2013.

\bibitem{HQNgo}
H.~Q. Ngo, E.~G. Larsson, and T.~L. Marzetta, ``Energy and spectral efficiency
  of very large multiuser {MIMO} systems,'' \emph{IEEE Trans. Commun.},
  vol.~61, no.~4, pp. 1436 -- 1449, Apr. 2013.

\bibitem{AMohammadi}
A.~Mohammadi and F.~M. Ghannouchi, ``Single {RF} front-end {MIMO}
  transceivers,'' \emph{IEEE Commun. Mag.}, vol.~49, no.~12, pp. 104--109, Dec.
  2011.

\bibitem{APitar}
A.~Pitarokoilis, S.~K. Mohammed, and E.~G. Larsson, ``On the optimality of
  single-carrier transmisson in large-scale antenna systems,'' \emph{IEEE
  Wireless Commun. Lett.}, vol.~1, no.~4, pp. 276--279, Apr. 2012.

\bibitem{YLiu2}
\BIBentryALTinterwordspacing
Y.~Liu, G.~Y. Li, Z.~Tan, and D.~Qiao. Signle-carrier modulation for
  large-scale antenna systems. [Online]. Available:
  \url{http://arxiv.org/abs/1508.00109.}
\BIBentrySTDinterwordspacing

\bibitem{DTse}
D.~Tse and P.~Viswanath, \emph{Fundamentals of Wireless Communication}.\hskip
  1em plus 0.5em minus 0.4em\relax Cambridge University Press, 2005.

\bibitem{YLiu}
Y.~Liu, Z.~Tan, and G.~Y. Li, ``Single-carrier modulation with {ML}
  equalization for large-scale antenna systems over {R}ician fading channels,''
  in \emph{Proc. IEEE Int. Conf. Acoust. Speech and Signal Process. (ICASSP)},
  2014, pp. 5759--5763.

\bibitem{AGoldsmith}
A.~Goldsmith, \emph{Wireless Communications (Chinese version)}.\hskip 1em plus
  0.5em minus 0.4em\relax Posts \& Telecom Press, 2007.

\bibitem{AKammoun}
A.~Kammoun, A.~M$\ddot{\text{u}}$ller, E.~Bj$\ddot{\text{o}}$rnson, and
  M.~Bebbah, ``Linear precoding based on polynomial expansion: Large-scale
  multi-cell {MIMO} systems,'' \emph{IEEE J. Sel. Topics Signal Process.},
  vol.~8, no.~5, pp. 861--875, Oct. 2014.

\bibitem{HHuh}
H.~Huh, A.~M. Tulino, and G.~Caire, ``Network {MIMO} with linear zero-forcing
  beamforming: Large system analysis, impact of channel estimation, and
  reduced-complexity scheduling,'' \emph{IEEE Trans. Inf. Theory}, vol.~58,
  no.~5, pp. 2911--2934, May 2012.

\bibitem{JJose2}
J.~Jose, A.~Ashikhmin, T.~L. Marzetta, and S.~Vishwanath, ``Pilot contamination
  and precoding in multi-cell {TDD} systems,'' \emph{IEEE Trans. Wireless
  Commun.}, vol.~10, no.~8, pp. 2640--2651, Aug. 2011.

\bibitem{ALiu}
A.~Liu and V.~K.~N. Lau, ``Two-stage subspace constrained precoding in massive
  {MIMO} cellular systems,'' \emph{IEEE Trans. Wireless Commun.}, vol.~14,
  no.~6, pp. 3271--3279, June 2015.

\bibitem{FSchaich}
F.~Schaich and T.~Wild, ``Waveform contender for 5{G}-{OFDM} vs. {FBMC} vs.
  {UFMC},'' in \emph{Proc. Connum., Control and Signal Process. (ISCCSP)},
  Athens, Greek, Feb. 2014, pp. 457--460.

\end{thebibliography}

\end{document}